# Ultrafast photocarrier recombination dynamics in black phosphorus-molybdenum disulfide (BP/MoS$_2$) heterostructure

*Zhonghui Nie[1], Yuhan Wang[1], Ziling Li[2], Yue Sun[1], Xiaoping Liu[3], Edmond Turcu[1], Yi Shi[1], Rong Zhang[1], Yu Ye[2], Yongbing Xu[1*], Fengqiu Wang[1*]*

[1]School of electronic science and engineering, Nanjing University, Nanjing 210023, China

[2]Department of physics, Peking University, Beijing 100871, China

[3]College of Engineering and Applied Sciences, Nanjing University, Nanjing 210093, China

**Abstract:**

Van der Waals (vdW) heterostructures constructed with two-dimensional (2D) materials have attracted great interests, due to their fascinating properties and potential for novel applications. While earlier efforts have advanced the understanding of the ultrafast cross-layer charge transfer process in 2D heterostructures, mechanisms for the interfacial photocarrier recombination remain, to a large extent, unclear. Here, we investigate a heterostructure comprised of black phosphorus (BP) and molybdenum disulfide (MoS$_2$), with a type-II band alignment. Interestingly, it is found that the photo-generated electrons in MoS$_2$ (transferred from BP) exhibit an ultrafast lifetime of ~5 ps, significantly shorter than those of the constituent materials. By corroborating with the relaxation of photo-excited holes in BP, it is revealed that the ultrafast time constant is as a result of efficient Langevin recombination, where the high hole mobility of BP facilitates a large recombination coefficient (~$2 \times 10^{-10}$ m$^2$/s). In addition, broadband transient absorption spectroscopy confirms that the hot electrons transferred to MoS$_2$ distribute over a broad energy range following an ultrafast thermalization. The rate of the interlayer Langevin recombination is found to exhibit no energy level dependence. Our findings provide essential insights into the fundamental photo-physics in type-II 2D heterostructures, and also provide useful guidelines for customizing photocarrier lifetimes of BP for high-speed photo-sensitive devices.

**INTRODUCTION:**

Two-dimensional (2D) materials, such as graphene,[1] transition metal dichalcogenides[2] (TMDs) and black phosphorus (BP),[3] exhibit a variety of unique properties, making them promising building blocks for future atomically-thin devices. Due to the weak interfacial interaction—van der Waals (vdW) forces, 2D materials can be easily fabricated into high-quality vdW heterostructures[4-5] through mechanical stacking without lattice matching constraints, offering a significant chance to explore exotic physics[6-8] and novel devices.[9-12] For example, despite different lattice structures of graphene and TMDs, graphene/TMDs heterostructures still exhibit robust interlayer coupling;[13] binding energy of excitons in the $WS_2$/graphene can be modulated by dielectric screening effect;[14] Coulomb interactions lead to long-lived interlayer excitons in monolayer $MoSe_2$/$WSe_2$ heterostructures.[15] These vdW heterostructures have also been studied for various device applications, especially optoelectronic devices, such as photodetectors,[11,12] solar cells.[16,17]

A comprehensive understanding of interlayer photocarrier behaviors and dynamics is the key to optimize the performance of optoelectronic devices based on vdW heterostructures.[11,12,16,17,23] Interlayer charge transfer, which plays a significant role in determining the light-matter interaction in vdW heterostructures, has received much attention recently. Thanks to the built-in field across the interface, photocarriers can easily transfer from one layer to another and this process has been identified to occur within tens of femtoseconds regardless of the twist angle between different layers.[13,18-22] However, thus far, studies on the behaviors of photocarriers after transferring into another layer remain scarce. Separation of electrons and holes changes both their distribution across the layers and recombination pathways (mostly non-radiative ones), which directly impacts the response speed and conversion efficiency of photo-sensitive devices. Recent observation of long-lived interlayer exciton emission in MoSe$_2$/WSe$_2$ heterostructure[15] and the remarkable elongation of photocarriers' lifetime in the tri-layer heterostructure through band alignment engineering[22] confirm the abundant photo-physics in the vdW heterostructures, but the investigations on the non-radiative relaxation mechanisms of interlayer photocarriers are still very limited. Fundamental aspects of photocarrier relaxation in adjacent layers, including the temporal dynamics as well as the associate physical mechanisms, remain poorly understood.

To provide answers to these open questions, type-II band alignment is required as it allows the probing of specific photocarriers (i.e. electrons or holes). We use BP/MoS$_2$ as an exemplary heterostructure, because BP and MoS$_2$ films can form type-II band structure (Fig.1(a)),[2,23,24] in the meantime mechanically exfoliated BP and MoS$_2$ are intrinsic *p*- and *n*-type semiconductors, which can form a building block for high-performance photonic devices. In addition, the direct and tunable bandgap of BP, spanning the visible to the mid-IR,[25-28] and the much-enhanced carrier mobility with regard to TMDs make BP a highly versatile candidate for 2D photonic devices.[23,25,26,28] The anisotropy of BP also provides a new degree of freedom to manipulate light-matter interactions.[25,26,29] Thus studying the BP/MoS$_2$ heterostructures not only promotes the understanding of fundamental properties of vdW heterostructures, but also facilitates the design of novel (mid-IR) devices.

In this study, ultrafast transient absorption spectroscopy is employed to investigate the photocarrier dynamics in the BP/MoS$_2$ heterostructure. In addition to observing electron transfer to MoS$_2$ after the photo-excitation of BP film, an unusual, ultrashort lifetime of ~5 ps for the separated photocarriers is identified. Such observation is attributed to strong Coulomb interaction of photocarriers across the layers, and can be well accounted for by considering the Langevin recombination model. The significant reduction of lifetime is ascribed to the enhanced Langevin constant caused by higher hole mobility in BP. Furthermore, broadband transient absorption spectroscopy is performed to reveal the energy distribution of the transferred hot electrons and their influence on the Langevin processes. Our findings contribute complementary insights into the physical mechanisms governing photocarriers dynamics in vdW heterostructures. The uncovered ultrafast non-radiative recombination pathway also provides guidance on designing high-speed optoelectronic and photovoltaic devices.

**RESULTS and DISCUSSION**

**Fabrication and characterizations of BP/MoS$_2$ heterostructure.** The BP and the MoS$_2$ films were mechanically exfoliated and stacked to form a heterostructure as shown in Fig. 1(a-b). To avoid degradation of the BP film, a thin boron nitride film was used to cover the whole heterostructure. The thickness of the BP and MoS$_2$ films are ~8 and 1.1 nm, respectively, as measured by the atomic force microscopy (AFM) shown in Fig. S1. The layer number of the BP film is estimated to be ~15.[29] The layer thickness value measured for the $MoS_2$ sample is close to the bilayer value (1.3 nm) and the difference of $E_{2g}^1$ and $A_{1g}$ Raman modes in $MoS_2$ is about 23 cm$^{-1}$, which indicate that the sample is bilayer $MoS_2$.[29,30] The characteristic Raman spectra of the BP film, bilayer $MoS_2$ and the heterostructure are presented in Fig. 1(c). These Raman modes associated with the individual materials have also been observed in the overlap region of the heterostructure, which is consistent with previous reports and suggests a good sample quality. In addition, because the $A_g^2$ mode in BP film originates from the in-plane atomic motion, its Raman intensity exhibits polarization dependence.[31] The crystal orientation of BP film can be easily identified through polarization-dependent Raman spectroscopy, as exhibited in Fig. S3. PL spectra of the bilayer $MoS_2$ and the heterostructure are presented in Fig. 1(d). The peaks at 680 nm and 620 nm of the $MoS_2$ (red line) arise from the A and B excitons.[2,30] Noticeable reduction (~30%) of PL intensity has been observed in the heterostructure, a typical signature for type-II heterostructures,[23,24] as shown in Fig. 1(a). The ~30% decrease is not significant compared with other TMDs-based heterostructures, and the main reasons could be the low PL efficiency of bilayer $MoS_2$ and the blocking of photo-excited holes from $MoS_2$ to BP film by the majority carriers in BP film (holes) with high density.[25,27]

**Photocarrier dynamics after charge transfer.** The ultrafast photocarrier response of the BP/MoS$_2$ heterostructure was directly monitored by transient differential reflection spectroscopy (details in Fig. S4). An 800 nm (1.55 eV) linearly polarized laser, with 100 fs pulse duration, worked as the pump beam to excite photo-carriers in the heterostructure. Due to the energy band arrangement as well as the layer thickness, pump absorption should be primarily from the BP film.[2,32] Using a 0.4%/nm absorption rate at 800 nm for BP film and assuming each photon absorbed excites one electron-hole pair, the peak density of photo-carriers excited in the BP can reach a density of ~10$^{12}$ cm$^{-2}$.[29] To observe the electron transfer from the conduction band (CB) of BP film to CB of MoS$_2$, we adopted a 620 nm laser beam (larger than the bandgap of MoS$_2$) as the probe, which can detect the lifetime of the transferred electrons in MoS$_2$, as illustrated in Fig. 2(a). Since the photon energy of the pump pulse (1.55 eV) is smaller than that of the probe pulse (2 eV), individual MoS$_2$ and BP film yielded no detectable signals, as shown in Fig. 2(b). However, an ultrafast and intense photo-bleaching signal has been observed in the heterostructure region, depicted by the black line in Fig. 2(b). While the steep rising feature (~400-500 fs) corresponds to the ultrafast cross-layer electron transfer,[18-21] more remarkable phenomenon is the extremely fast relaxation time of ~5±0.2 ps (as fitted by single-exponential function). Such a short lifetime is not only at least one order of magnitude faster than the intrinsic lifetimes of photocarriers in individual MoS$_2$ and BP film, as shown in Fig. S5,[29,33-35] but also strikingly different from previous reports on TMD heterostructures.[18,21] To explore the origin of such ultrafast lifetime, power-dependent measurements have been carried out, as exhibited in Fig. S6. We extracted the peak values and lifetimes from the dynamic curves and summarized them as a function of pump fluence, as illustrated in Fig. 2(b). Clearly, below the fluence value of ~80 μJ/cm$^2$ (or photocarrier density of ~10$^{13}$ cm$^{-2}$), the transient signal increase linearly with the incident power.[29,33] Interestingly, the lifetime of the transferred electrons in MoS$_2$ stays almost constant (~5 ps), and exhibits no clear dependence on photocarrier density, even when the density has gone beyond the linear region, indicating that such short electron lifetime should not arise from ultrafast trapping from interfacial defects.[36]

To reveal more details of the relaxation mechanisms, we changed the probe wavelength to 1050 nm when the pump beam was fixed at 800 nm. The probe pulse now primarily detects transient signal from holes in the BP film, due to its small photon energy (~ 1.2 eV), far below the bandgap of MoS$_2$, and the schematic diagram is shown in Fig. 2(a). The measured transient signals of the BP film and the heterostructure under 800 nm pump and 1050 nm probe are presented in Fig. 2(d). The transient signal of individual BP film can be well fitted by a time constant of 130 ps (see Fig. S5). Due to the weak exciton binding energy in the BP film,[36] we attribute such a process to the lifetime of free carries in BP film and similar result has also been reported before.[29,33,35] Noticeably, a fast and enhanced relaxation component (yellow shade area) clearly emerges in the dynamic curve of the BP/MoS$_2$ heterostructure, where the magnitude of the signal is 3-4 times larger than that of the individual BP, and the initial decay component is much faster, as described by a time constant of ~6±0.5 ps. Such an enhancement has been tentatively ascribed to the increase of screening effect and similar phenomenon has also been found in graphene/WS$_2$ heterostructure.[13] It should be noted that this first relaxation time of ~6±0.5 ps, corresponding to the ultrafast reduction of holes' population in the BP film, is consistent with the lifetime of the transferred electrons, as measured in Fig. 2(b). The coincidence between lifetime of electrons in MoS$_2$ and that of holes in BP film strongly suggests that there may exist an efficient bimolecular recombination process for interlayer electrons and holes. In the vdW heterostructures, there are two mechanisms potentially contributing to this unusual interlayer *e-h* recombination: Shockley-Read-Hall (SRH) recombination, assisted by traps or defects states in the gap; and Langevin recombination, dominated by interlayer Coulomb interaction, as illustrated in Fig. 3(a).[38-41] The trap-assisted SRH recombination can be ruled out as the dominant process by the observed fluence-independent results. The Langevin model describes the simultaneous recombination between electrons and holes by Coulomb interaction, which is consistent with our observations in Fig. 2. The Langevin model could be applicable here with another important reason being the enhanced Coulomb interaction between photocarriers confined in low-dimensional systems.

The Langevin recombination rate *R* can be expressed as:[38,39]

$$R = B \times (n_0 + \Delta n) \times (p_0 + \Delta p)^s$$

where $B$ is the Langevin recombination constant, $n_0$ and $p_0$ are respectively the intrinsic density of electrons in MoS$_2$ and holes in BP, and $\Delta n$ and $\Delta p$ are photo-generated electron and hole density. An exponent $s$ (1.2<$s$<1.5) is typically used for the 2D scenario.[40] As is shown by Fig. 3(b), the relationship between the observed lifetime of electrons in MoS$_2$ and pump fluence can be well fitted by the Langevin model, when a typical value of 1.2 is adopted for $s$. The Langevin recombination constant $B$ was extracted by the fitting to be ~2 × 10$^{-10}$ m$^2$/s, larger than what has been reported by Chul-Ho Lee *et. al* for a MoS$_2$/WSe$_2$ heterostructure ($B$~ 4.0 × 10$^{-13}$ m$^2$/s and $s$=1.2).[38,40] The larger $B$ in our case could be due to the much higher mobility of BP film.[25,27] (Detailed fitting procedures and discussion can be found in Supplementary Note 2) Therefore, it is reasonable to assign the observed ultrashort lifetime of photocarriers to an efficient Langevin recombination process with a large recombination constant $B$, due to the high mobility of holes in BP film. This may provide a useful way for tuning the recombination rate or photocarrier lifetime of 2D heterostructure.

**Broadband ultrafast spectroscopy.** We then turn to illustrate the detailed temporal dynamics and energy distribution of the electrons after making the cross-layer transfer. The probe wavelength was tuned from 550 nm (2.25 eV) to 675 nm (1.84 eV), covering the A and B excitonic peaks of MoS$_2$, while the pump beam was fixed at 800 nm. The corresponding broadband transient spectroscopies of BP/MoS$_2$ heterostructure is shown in Fig. 4(a), and the red lines are single-exponential fittings.

First, the photo-bleaching signals in Fig.4(a) suggest that the electrons occupy quite high energy states with a broad spread in the CB of $MoS_2$. This observation agrees with some recent theoretical and experimental results, where hot electrons are able to transfer from CBM of one layer to conduction bands above the CBM of the other layer.[43,44] Second, the excess energy available to the transferred electrons could induce thermalization process. To evaluate the thermalization time constant, the peak values of the transient signals in Fig. 4(a) is extracted and plotted as a function of the probe wavelengths in Fig. 4(b). The trend of peak values roughly follows that of the PL spectrum of $MoS_2$ (the black solid line), confirming that thermal distribution of electrons is nearly established.[45] These observations mean that the ultrashort rising time (300-400 fs) in Fig. 4(a) correspond to both the electron transfer and thermalization processes. Then, we examined the relationship between electron lifetime and the probe wavelength, as shown in Fig. 4(c). Interestingly, the lifetime varying between 5 and 8 ps shows no obvious dependence on the probe wavelength, which suggests that the rate of interlayer Langevin recombination is not sensitive to energy of electrons. Such dynamics is also interesting in that it suggests the cross-layer Langevin recombination in this $BP/MoS_2$ heterostructure does not have to be a band-edge process. Besides the energy distribution, we also investigated the polarization dependence of the transferred electrons and found that the anisotropy of photocarriers in the heterostructure has been partially weakened, compared with that of individual BP film (Supplementary Note 3).

Combining all our measurements, the transient photocarrier process in the BP/MoS$_2$ heterostructure can be characterized by three major steps: the generation of hot electron-hole pairs in BP through optical excitation; ultrafast interlayer charge separation (the electrons transfer to the MoS$_2$ layers while the holes are left in the BP layers) and thermalization (re-distribute within the conduction band of MoS$_2$) within few hundreds of femtoseconds, and finally the Langevin recombination between the transferred electrons and holes in different layers on a timescale of ~5 ps. These processes are schematically illustrated in Fig. 4(d). The response time of BP-based photodetector is generally limited by the intrinsic lifetime of photocarriers (up to hundreds of picoseconds) and such a significant reduction of photocarrier lifetime in the BP/MoS$_2$ heterostructure, close to that of graphene, makes it possible to construct BP heterostructures to realize novel IR photo-sensitive devices with high speed.

**CONCLUSIONS**

In conclusion, we have investigated the ultrafast dynamics of photocarriers in the BP/MoS$_2$ heterostructure by optical two-color ultrafast pump-probe spectroscopy. In addition to observing ultrafast charge transfer, we found an unusual, ultrashort lifetime of ~5 ps, significantly shorter than those of the individual constituent materials. Combined with theoretical analysis, we attributed the observed ultrafast relaxation of interlayer electrons and holes to the Langevin recombination, with an enhanced Langevin constant *B*, due to the much higher hole mobility in BP. Moreover, broadband measurements reveal a relatively broad energy distribution of transferred electrons in $MoS_2$ which exhibits no obvious influence on the rate of the subsequent Langevin recombination, suggesting this interlayer recombination in the $BP/MoS_2$ heterostructure is not necessarily a band-edge process. Our findings provide new insights into the understanding of photo-physics in vdW heterostructures and point out a new route to dramatically shorten photocarrier lifetimes at 2D heterointerface through enhanced Langevin process, which is relevant for developing ultrafast optoelectronic devices.

ASSOCIATED CONTENT

**Supporting Information.**

AFM of the heterostructure, Raman spectrum of $MoS_2$, anisotropic intensity of Raman modes in BP, optical layout of setup, transient signal of BP, dynamics of the heterostructure under different fluences, and Langevin model fitting and the anisotropic dynamics of the heterostructure.

Supplementary Note 1: Pump-probe setup

Supplementary Note 2: Langevin recombination model

Supplementary Note 3: Anisotropic dynamics of BP/$MoS_2$ heterostructure

**Corresponding Author**

*E-mail: fwang@nju.edu.cn

*E-mail: ybxu@nju.edu.cn

**Author Contributions**

F.W. and Y.X. conceived the project. Z.N. and Y.W. performed the ultrafast measurements and basic characterizations of the samples and Z.L. and Y.Y. prepared the samples. The manuscript was written through contributions of all authors. All authors have given approval to the final version of the manuscript.

**ACKNOWLEDGMENT**

**Figures:**

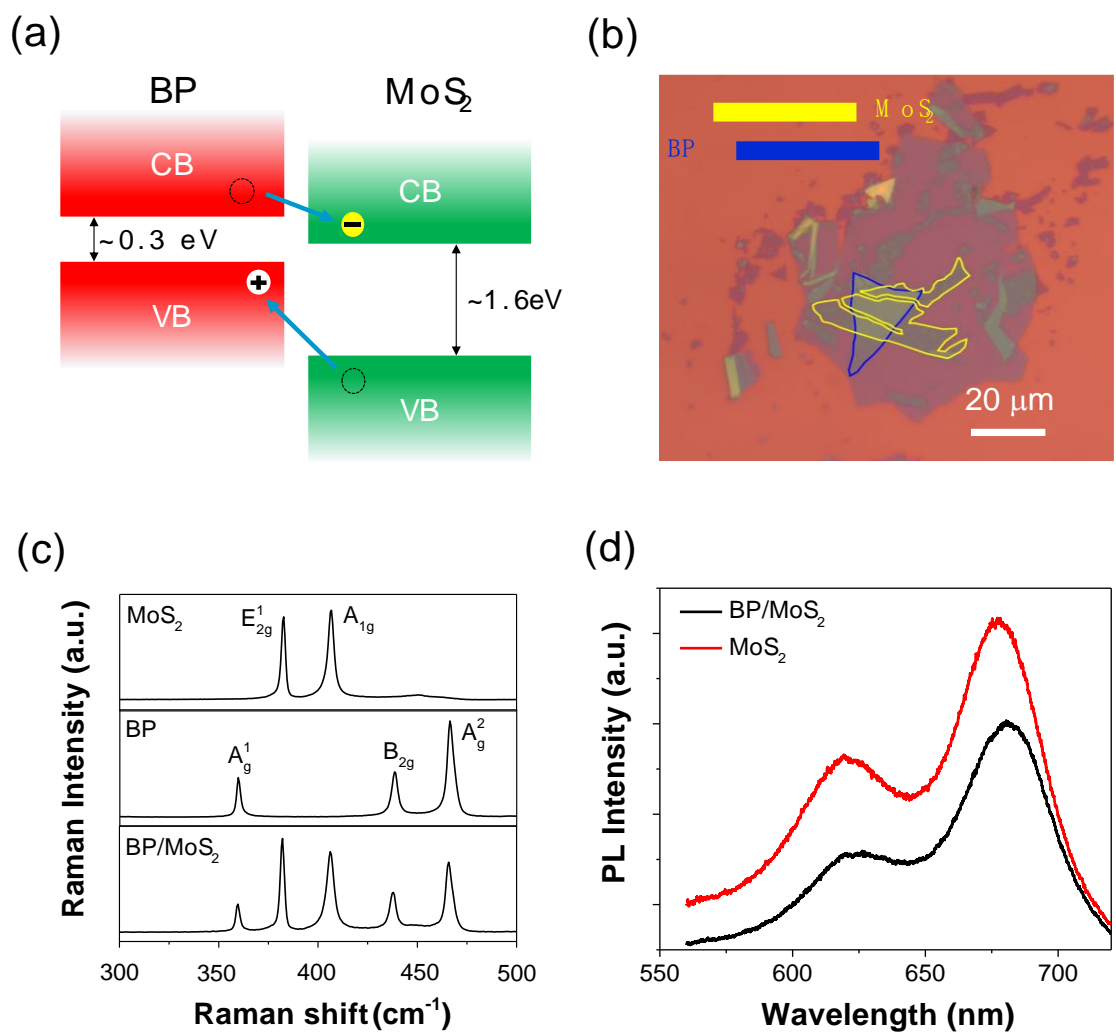

**Fig. 1** (a) Illustration of the band alignment of BP/MoS$_2$ heterostructure, belonging to Type-II heterostructure. Such band structure can easily separate electrons and holes into different layers. (b) Optical picture of BP/MoS$_2$ heterostructure, where blue line represents BP film and yellow line corresponds to bilayer MoS$_2$. The scale bar is 20 μm. It should be noted that a thin hBN film (not marked) was used to protect BP film from degradation. (c) Characterized Raman spectra of MoS$_2$, BP and its heterostructure. (d) Photoluminescence spectra of MoS$_2$ and the heterostructure under 512 nm laser excitation. PL quenching has been observed, which is attributed to interlayer charge transfer in the heterostructure.

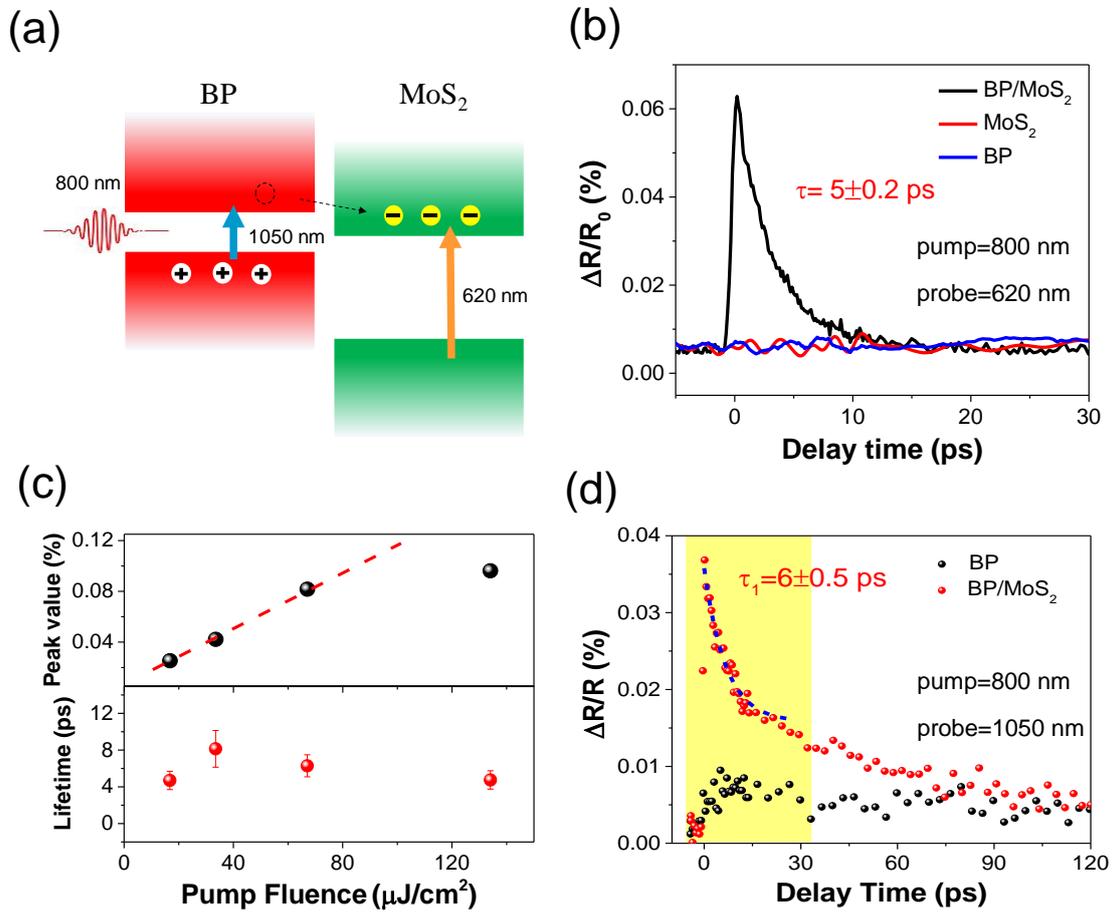

**Fig. 2** (a) Schematic describing the transient process when probe wavelengths are 620 nm and 1050 nm. (b) Transient signals of the heterostructure and MoS$_2$ respectively when 800 nm works as pump and 620 nm as probe beam. The ultrafast signal proves the existence of electron transfer and the relaxation time is only ~ 5 ps, much shorter than those of photo-carriers in individual MoS$_2$ and BP. (c) The summary of peak values and lifetimes, extracted from (a), plotted as functions of pump fluence. (d) Time-resolved differential reflection measured from BP film and BP/MoS$_2$ heterostructure with a 800 nm pump and a 1050 nm probe.

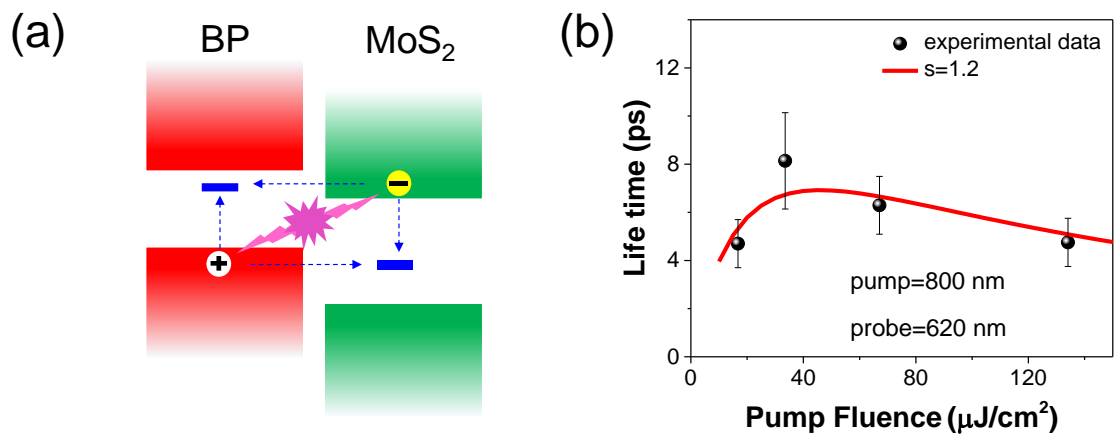

**Fig. 3** (a) Two potential interlayer recombination mechanisms. Blue and pink arrows indicate SRH and Langevin recombination processes, respectively. The blue solid lines located in the bandgap of materials represent traps or defects. (b) The lifetimes under different pump fluences, extracted from Fig. 3(b), and blue and red lines are fitting results of the Langevin model when s=1.2.

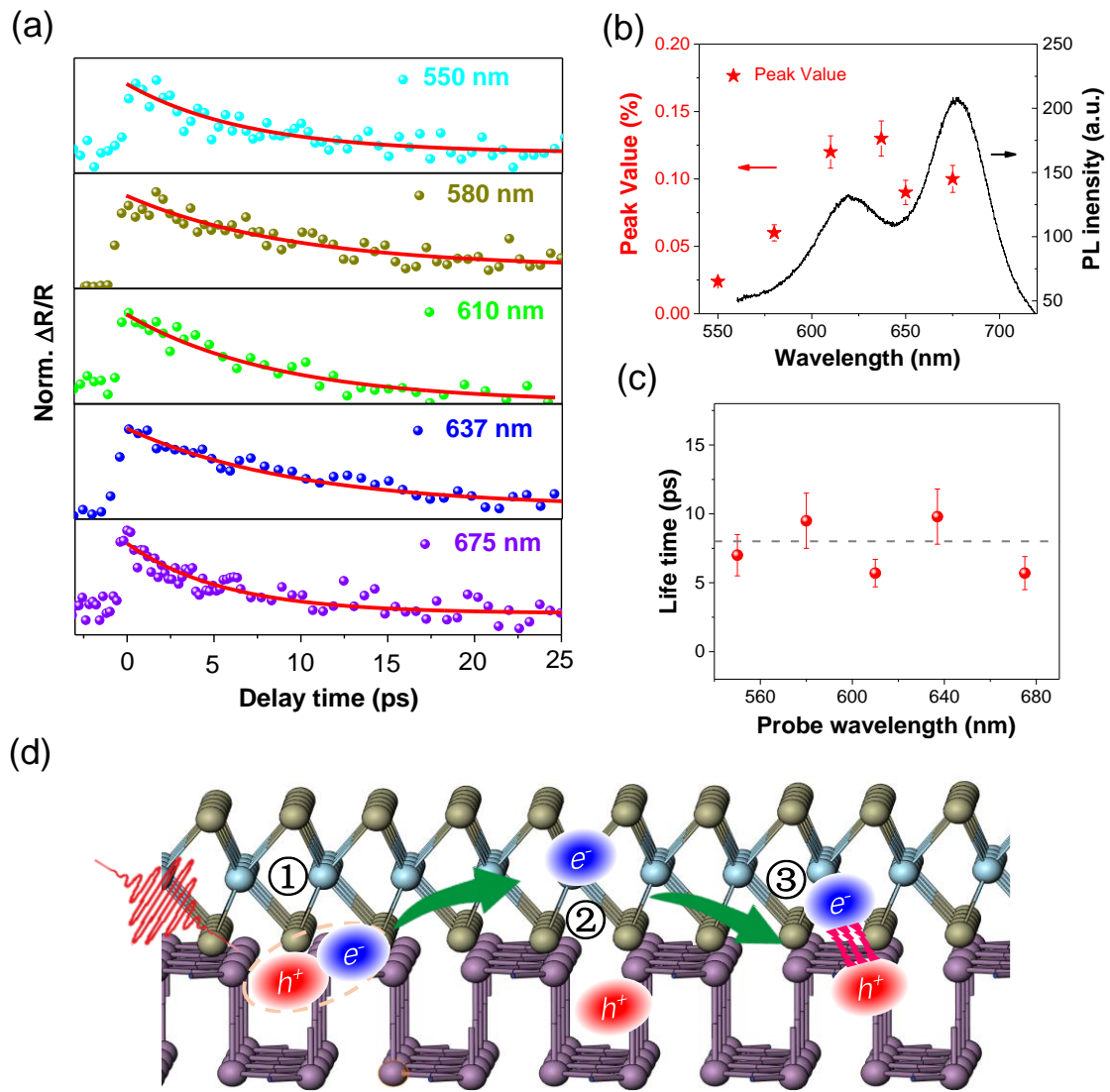

**Fig. 4** Broadband ultrafast nonlinear optical properties of BP/MoS$_2$ heterostructure. (a) The ultrafast pump-probe measurements with probe wavelength varying from 550 nm to 675 nm, including A and B exciton peaks, when pump wavelength is fixed at 800 nm. Coloured dots are experimental data and solid lines correspond to single-exponential fitting results. (b) Peak values of dynamic curves in the heterostructure under different probe wavelengths (red stars, left axis), extracted from (a). For comparison, the PL spectrum (black line, right axis) of bilayer MoS$_2$ has been also exhibited, suggesting that exciton states in $MoS_2$ may affect peak values of transferred electrons. (c) The fitted lifetime as a function of probe wavelength, extracted from (a). The lifetime of the heterostructure seems constant when changing probe wavelength. (d) Illustration of the dynamics in the heterostructure. After excitation, ① electron-hole pairs are generated in BP film, and then ② hot electrons transfer to $MoS_2$, due to the potential drop and go through the thermalization and re-distribution processes. Finally ③ electrons and holes at different layers gradually recombine within very short time ~5-8 ps through efficient Langevin recombination.